\documentclass[amsmath,amssymb,aps,pra,twocolumn]{revtex4-2}
\usepackage{graphicx}
\usepackage{dcolumn}
\usepackage{bm}
\usepackage{hyperref}
\usepackage[mathlines]{lineno}
\usepackage{lineno}
\usepackage{graphicx}
\usepackage{amsfonts} 
\usepackage{dsfont}
\usepackage{braket} 
\usepackage{hyperref} 
\usepackage{xcolor}

\begin{document}

\preprint{APS/123-QED}

\title{Bell non-locality in two-mode Gaussian states revealed via local squeezing}

\author{A. Lezama}
\email{alezama@fing.edu.uy}
\author{A. Auyuanet}
\affiliation{Instituto de F\'{\i}sica, Facultad de Ingenier\'{\i}a,
Universidad de la Rep\'{u}blica,\\ J. Herrera y Reissig 565, 11300
Montevideo, Uruguay}

\date{\today}

\begin{abstract}
Local unitary operations cannot affect the quantum correlations between two systems sharing an entangled state although they do influence the outcomes of local measurements. By considering local squeezing operations we introduce an extended family of observables allowing violation of the CHSH Bell inequality for two-mode Gaussian systems. We show that local squeezing can enable or enhance the identification of non-local two-mode states. In particular, we show that local squeezing followed by photons/no-photons discrimination can suffice to reveal non-locality in a broad ensemble of pure and mixed two-mode Gaussian states. 
\end{abstract}

\maketitle


\section{Introduction}
Non-locality occurs while the correlations between measurement outcomes performed in two systems (Alice's and Bob's) cannot be explained as the result of random processes driven by  stochastic variables (the so-called hidden variables) described by local probability distributions. 

Non-locality is the strongest quantum correlation between two-systems. It implies bidirectional steering and entanglement \cite{Wiseman07,Brunner14}. In his seminal work, Bell has derived inequalities that all hidden variable processes must verify \cite{Bell64}. In consequence, violation of a Bell inequality implies Bell non-locality (BNL).\\

A convenient form of Bell inequality suitable for binary output measurements was introduced by Clauser, Horne, Simony and Holt (CHSH) \cite{Clauser69}. It states:
\begin{eqnarray}\label{CHSH}
\mathcal{B}_{CHSH}\equiv\vert\braket{AB}+\braket{AB^{\prime}}+\braket{A^{\prime}B}-\braket{A^{\prime}B^{\prime}} \vert \leq 2
\end{eqnarray}
where $\{ A,A^{\prime}\}$ and $\{ B,B^{\prime}\}$ are pairs of observables (also designated as setups), acting on Alice's and Bob's systems respectively, whose possible outcomes are $+1$ and $-1$. The maximum violation of the CHSH inequality allowed by Quantum Mechanics is $\mathcal{B}_{CHSH}  = 2\sqrt{2} \simeq 2.83$ \cite{Cirel80}. 
After extensive theoretical advances and sophisticated experimental tests, BNL remains a topic of large current interest \cite{Thearle18,Kitzinger21,loopholeNature}. \\

Most initial work regarding BNL was concerned with low dimensional systems \cite{Brunner14}. Regarding continuous variable systems, a strong argument against the possibility of BNL in Gaussian states was raised by Bell \cite{Bell87}. In essence the argument states that since Gaussian states are represented in phase-space by positive Gaussian quasi-probability distributions (one example being the Wigner function), such distributions could effectively behave as a probability distribution for hidden variables (here complex numbers designating position in phase-space) determining the results of possible measurement outcomes. \\

However, an important detail seem to have escaped the attention of Bell. In order for this argument to hold it is necessary that the function relating the measurement outcome to the hidden variable is bounded by the maximum and minimum outcome values. While this is true in classical physics, it is not necessarily the case in quantum physics. A paradigmatic example is the parity operator $\Pi = (-1)^{\hat{n}}$ where $\hat{n}$ is the number operator. Its possible outcomes are $\pm 1$ however its  Wigner representation is $\pi\delta(\alpha)/2$, where $\delta(\alpha)$ is the complex Dirac's delta function which is unbounded \cite{Royer77,Banaszek98,Praxmeyer05}.\\

The possibility to demonstrate non locality in Gaussian states was first suggested by Grangier, Potasek, and Yurke (GPY)\cite{Grangier88} considering a scenario were the fields received by Alice and Bob are first displaced in phase-space by mixing with coherent states in a beamsplitter and then detected. A CHSH Bell inequality was constructed based on coincidence and single counts of the two detectors. It was also shown that a two-mode squeezed vacuum state (TMSV), also known as generalized EPR states,  would violate such inequality.\\ 

Banaszek and W\'odkiewicz (BW) \cite{Banaszek98} were first to establish that measurements of the displaced parity operator can lead to a CHSH Bell inequality violation in the case of TMSVs. In their derivation BW assumed that only one of the two displacements performed by Alice or Bob was nonzero. Under such assumption (often designated as BW maximization scheme) they have shown that a maximum value of $\mathcal{B}_{CHSH}=2.19$ \cite{Ferraro05} can be reached in the case of an infinitely squeezed TMSV (EPR state).\\

It is interesting to observe that the experimental scheme suggested in GPY only required detectors able to discriminate between the presence and absence of photons (a realistic assumption in view of the  detectors available at that time). On the other hand, measurement of the parity operator as in BW requires photon-number resolving detectors, not available until recently.\\ 

Several works have extended the initial suggestions of GPY and BW either by generalizing the BW maximization scheme through the consideration of all four phase-space displacements (two performed by Alice and two by Bob) \cite{Jeong03,Lee09} or through the introduction of new observables allowing CHSH-Bell inequality violations for Gaussian states \cite{Chen02,Jeong03,Praxmeyer05}.

It was established that measurements of the displaced parity operator or the displaced presence/absence of photons cannot lead to a maximal violation of the CHSH inequality, the largest values of $\mathcal{B}_{CHSH}$ being 2.32 and 2.45 respectively \cite{Jeong03,Lee09}. However, maximal violation of the CHSH inequality in Gaussian states is possible with more complex  (pseudo-spin) observables obeying  angular momentum commutation rules \cite{Chen02,Jeong03,Praxmeyer05}.\\ 

The first experimental demonstration of Bell non-locality in two-mode Gaussian states is due to Kuzmich and coworkers \cite{Kuzmich00} essentially following the experimental scheme suggested by GPY \cite{Grangier88}. Evidence of violation of the CHSH inequality consistent with the upper limit $\mathcal{B}_{CHSH} =2.45$ was presented.\\

The possibility to violate CHSH Bell inequalities in Gaussian states using binary outcomes operators was extensively analyzed in \cite{Lee09}. Generalizing the proposals in GPY and BW, a broad family of observables $\hat{O}(\alpha;s)$ depending on a complex variable $\alpha$ (corresponding to a displacement in phase space) and parameterized by a non-positive real number $s$ was introduced. Each of these observables has an unbounded representation in the corresponding $s$ parameterized (Gaussian) quasi-probability distribution (sQPD) \cite{Gerry05} although the maximum and minimum possible measurement outcomes are always $\pm 1$. 

As shown in \cite{Lee09}, for each of these observables a CHSH Bell inequality can be derived where the required setups correspond to two different choices of displacements ($\alpha$, $\alpha^{\prime}$) for Alice and ($\beta$, $\beta^{\prime}$) for Bob. In practice, the necessary experimental setups involve 
combining the corresponding field mode with a coherent state local oscillator, followed by light detection using a photon number resolving detector \cite{Banaszek99}. An important exception concerns the case $s=-1$ for which the detector  is only required to discriminate the vacuum from non-vacuum states. The observables considered in BW (displaced parity) or in GPY correspond to the particular cases $s=0$ and $s=-1$ respectively.\\

In the previously mentioned work only phase-space displacements were consider to be applied by Alice or Bob prior to detection. However, it was noticed in \cite{Stobinska07} that nonlinear local (Kerr) interactions could lead to a significant increase in the CHSH inequality violation in the case of entangled two-mode coherent states. Following this result, Paternostro et al. \cite{Paternostro09} considered the Gaussian states produced by dividing single-mode squeezed state in a balanced beamsplitter which can be expressed as linear combination of two-mode entangled coherent states. They have shown that nonlinear Kerr interactions can assist to reveal CHSH inequality violation in such states. They also noticed that since such states can be obtained from TMSV states by local squeezing operations, the considered scheme could be used to reveal non-locality in TMSV states. More recently, Kitzinger et al. \citep{Kitzinger21} considered the use of local squeezing operations in a proposal for the observation of BNL between two Bose-Einstein condensates. Local squeezing operations were also included in an automated protocol designed for device-independent quantum key distribution \citep{Valcarce23}.\\

\begin{figure}[ht!]
\centering
\includegraphics[width=1\linewidth]{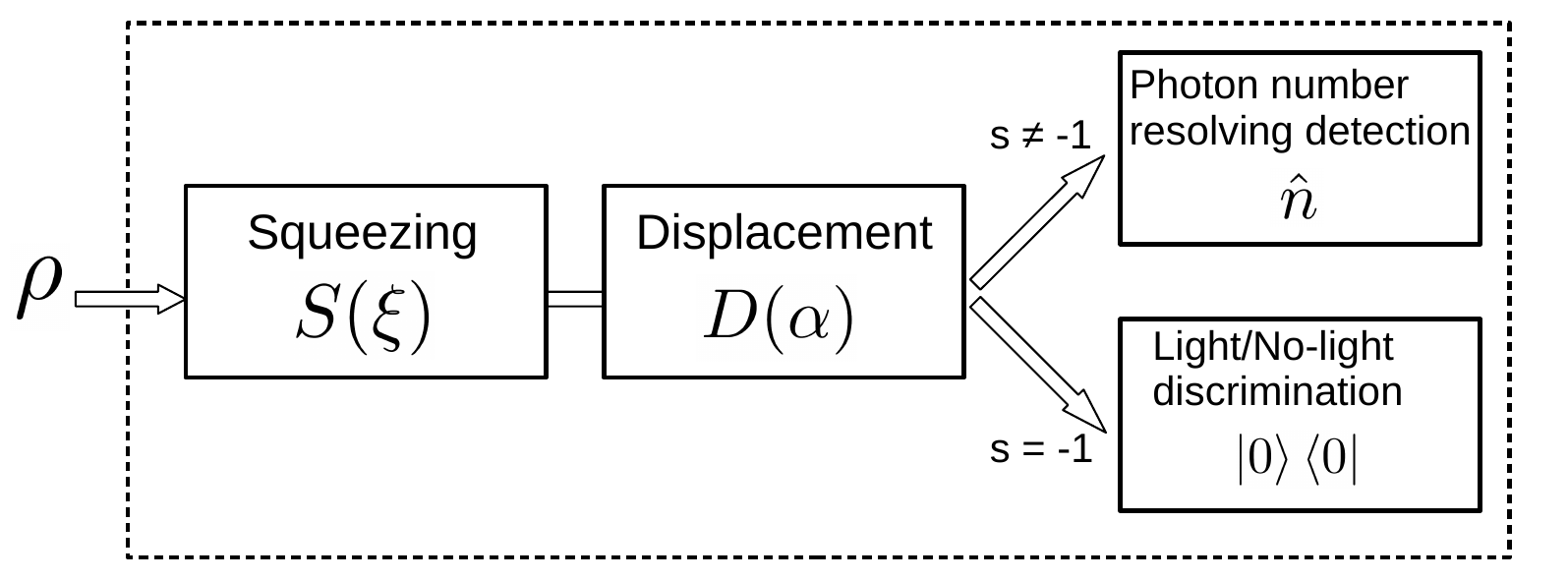}
\caption{\label{setup} Schematic representation of the setup corresponding to the implementation of the extended family of observables introduced in this article. Complementing previous schemes, the setup includes a local squeezing operation.}
\end{figure}

The aim of the present work is to present an extension of the set of observables considered in \cite{Lee09} by including local squeezing  operations $S(\xi)=\exp\left[\frac{1}{2}(\xi^{*}a^2-\xi a^{\dagger 2})\right]$  where $a$ and $a^{\dagger}$ are the annihilation and creation operators and $\xi$ a complex parameter. The setups corresponding to the extended family of observables $\hat{O}(\xi;\alpha;s)$ are schematically represented in Fig. \ref{setup}. The new observables depending  on two independent complex parameter $\xi$ and $\alpha$ lead to a generalization of the CHSH Bell inequalities. We show that such generalization increases the ensemble of two-mode Gaussian states for which BNL can be proven. Moreover, we show that for some states, squeezing-only observables  $\hat{O}(\xi,0,s)$ where the displacement  parameters $\alpha$ and $\beta$ are zero, are sufficient to reveal BNL. \\

\section{Extended observables family}\label{extendedObs}

We begin with a reminder of the $s$-parameterized observables $\{\hat{O}(\alpha;s)\}$ presented in \cite{Lee09} and their connection with the corresponding $s$-parameterized quasi-distributions.\\

The building blocks of the single-mode observables $\{\hat{O}(\alpha;s)\}$ are the projectors $\ket{\alpha,n}\bra{\alpha,n}$ onto the displaced photon number states $\ket{\alpha,n}=D(\alpha)\ket{n}$ where $D(\alpha)\equiv \exp(\alpha a^{\dagger}-\alpha^{*}a)$ is the displacement operator and $\ket{n}$ is a Fock state.\\ 

We directly extend the family $\{\hat{O}(\alpha;s)\}$ to $\{\hat{O}(\xi;\alpha;s)\}$ by considering the broader ensemble of projectors onto the \emph{squeezed} displaced number states $\ket{\xi,\alpha,n}=S(\xi)D(\alpha)\ket{n}$. In consequence the observables $\Pi(\alpha;s)$ introduced in \cite{Lee09} are replaced by:
\begin{eqnarray}
\hat{\Pi}(\xi;\alpha;s) &=& \sum_{n=0}^{\infty}\left( \frac{s+1}{s-1}\right)^n\ket{\xi,\alpha,n}\bra{\xi,\alpha;n}\label{nuevoPi}
\end{eqnarray}
from which the observables $\hat{O}(\xi;\alpha;s)$ are defined as in Eq. (3) of \cite{Lee09}:
\begin{equation}
\hat{O}(\xi;\alpha;s)=\begin{cases}
    (1-s)\hat{\Pi}(\xi;\alpha;s)+s\mathds{1}, & \text{if $-1 < s \leq 0$}.\\
    2\hat{\Pi}(\xi;\alpha;s)-\mathds{1}, & \text{if $ s \leq -1$}.
  \end{cases}
\end{equation}

By construction, the states $\ket{\xi,\alpha,n}$ are eigenstates of  
$\hat{\Pi}(\xi;\alpha;s)$ and the corresponding eigenvalues are $\left( \frac{s+1}{s-1}\right)^n$. As a result, the maximum and minimum outcomes of observables $\hat{O}(\xi;\alpha;s)$ are $\pm 1$.\\

Considering the joint observables $\hat{O}_A(\xi_a,\alpha;s)\otimes \hat{O}_B(\xi_b,\beta;s)$ acting on two systems $A$ and $B$ results in the CHSH bell inequality: 
\begin{align}
&\vert\langle\hat{O}(\chi_a;s)\otimes \hat{O}(\chi_b;s)+\hat{O}(\chi_a;s)\otimes \hat{O}(\chi^{\prime}_b;s) \nonumber\\
& +\hat{O}(\chi^{\prime}_a;s)\otimes \hat{O}(\chi_b;s)-\hat{O}(\chi^{\prime}_a;s)\otimes \hat{O}(\chi^{\prime}_b;s) \rangle\vert \leq 2
\end{align}
where $\chi_a$ and $\chi_b$ are shorthand notations for $\{\xi_a;\alpha\}$ or $\{\xi_b;\beta\}$.\\

The operators defined in Eq. \eqref{nuevoPi} are directly linked to an extension of the family of sQPD. We will refer to the functions in the extended family as squeezing-extended quasi-probability distributions SEQPD.  For a single mode the $s$ parameterized SEQPD is defined as:
\begin{subequations}
\begin{eqnarray}
W(\xi;\alpha;s)&=& \frac{2}{\pi(1-s)}Tr[\hat{\rho}\hat{\Pi}(\xi;\alpha;s)]\label{WseneralizadaDef}\\
& = & \frac{2}{\pi(1-s)}Tr[ S(\xi)^{\dagger}\hat{\rho}S(\xi)\hat{\Pi}(\alpha;s)] \label{Wsg_enfuncion_de_antigua}
\end{eqnarray}
\end{subequations}
where $\hat{\Pi}(\alpha;s)\equiv \hat{\Pi}(0,\alpha;s)$ is the operator appearing in the definition of the usual sQPD $W(\alpha;s)$ \cite{Lee09}.\\

From \eqref{Wsg_enfuncion_de_antigua} it can be seen that a SEQPD corresponds to the usual sQPD evaluated \emph{after} a squeezing operation on the state (see Fig. \ref{setup}). Since a squeezed Gaussian state remains a Gaussian state, it results that the extended quasi-probability distributions of Gaussian states are also Gaussian functions in phase-space.\\

\begin{figure}[ht!]
\centering
\includegraphics[width=1\linewidth]{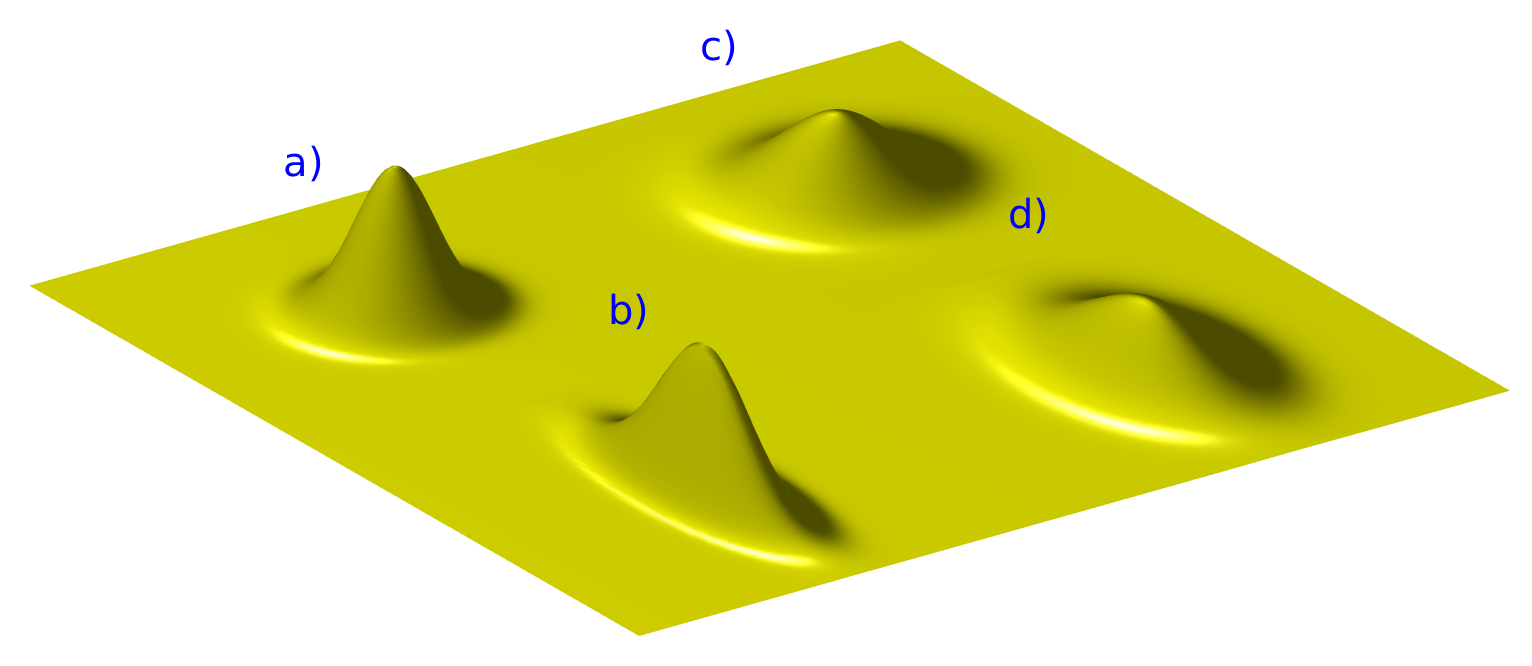}
\caption{\label{cuatro} Surface representation of four extended quasiprobabiliy distributions $W(u,\alpha;s)$ for the vacuum ($u \equiv \exp(2\xi)$). a) $W(u=1,\alpha;s=0)$ is the Wigner function. b) $W(u=3,\alpha;s=0)$, the stretched Wigner function resulting from squeezing. c) $W(u=1,\alpha;s=-1)$, the Husimi Q function. d) $W(u=3,\alpha;s=-1)$, Q function after squeezing operation. }
\end{figure}

As an illustration of SEQPDs, Fig. \ref{cuatro} depicts four different representation of the vacuum. a) and c) are the sQPD  $W(\alpha;s)$ for $s=0$ (Wigner function) and $s=-1$ (Husimi Q function) respectively. b) and d) correspond to the SEQPD  $W(\xi;\alpha;s=0)$ and $W(\xi;\alpha;s=-1)$ for $\exp(2\xi)=3$. 

It is worth noticing that, in consistence with Eq. \eqref{solowigner}, for $s=0$ the SEQPD (b) can be generated from the sQPD (a) through changes in scale along two orthogonal axis by factors with unit product. Also the maximum values of the two representations are equal.  The same does not occur for $s=-1$ [(c) and (d)]. In particular the maximum values of the representations in (c) and (d) are different. As will be shown below, the change in the maximum value of the sQPD evaluated after squeezing conveys information which can be used to demonstrate non-locality.\\

The CHSH Bell inequalities derived in \cite{Lee09} can be immediately generalized under the substitution of $W(\alpha;\beta;s)$ by $W(\chi_a;\chi_b;s)$ They are:
\begin{subequations}\label{CHSHChi}
\begin{align}
\mathcal{B}_{CHSH}\equiv\vert \frac{\pi^2(1-s)^4}{4}\mathds{C}(\chi_a,\chi_a^{\prime},\chi_b,\chi_b^{\prime})& \nonumber\\
 +\pi s(1-s)^2[W(\chi_a,s)+W(\chi_b,s)]+2s^2\vert &\leq 2 
\end{align} 
if $-1 < s \leq 0$ or 
\begin{align}
\mathcal{B}_{CHSH}\equiv\vert \frac{\pi^2(1-s)^2}{4}\mathds{C}(\chi_a,\chi_a^{\prime},\chi_b,\chi_b^{\prime})& \nonumber\\
-2\pi (1-s)[W(\chi_a,s)+W(\chi_b,s)]+2\vert &\leq 2 
\end{align} 
\end{subequations}
if $ s \leq -1$, where
\begin{align}
& \mathds{C}(\chi_a,\chi_a^{\prime},\chi_b,\chi_b^{\prime}) \equiv W(\chi_a,\chi_b,s)\nonumber\\
& +W(\chi_a,\chi_b^{\prime},s)+W(\chi_a^{\prime},\chi_b,s)-W(\chi_a^{\prime},\chi_b^{\prime},s)
\end{align}  

We next assume that Alice and Bob share a two-mode Gaussian state whose covariance matrix is given in the standard form \citep{Duan00,Adesso14};
\begin{eqnarray}\label{StandardI}
\sigma &= &  \begin{pmatrix} p & 0 & m & 0\\ 
0 & p & 0 & n\\
m & 0 & q & 0\\
0 & n & 0 & q
\end{pmatrix}
\end{eqnarray}

Alice and Bob may perform local squeezing operations. We consider only real squeezing parameters and introduce the notation $u\equiv \exp(2\xi_a)$ and $v\equiv \exp(2\xi_b)$. The corresponding symplectic transform $\mathcal{S}$ will be referred to as $\mathcal{S}_{uv}$ We have: 
\begin{equation}\label{squeezedStandardI}
\mathcal{S}_{uv} = Diag(u^{\frac{1}{2}},u^{-\frac{1}{2}},v^{\frac{1}{2}},v^{-\frac{1}{2}})
\end{equation} 

In the following we refer to $W(\chi_a;\chi_b;s)$ as $W(u;v;\alpha;\beta;s)$. In particular, $W(1;1;\alpha;\beta;s)$ designates the sQPD for a covariance matrix in the standard form.\\ 

The details of the evaluation of $W(u;v;\alpha;\beta;s)$ are described in the Appendix. \\ 

\section{Results} \label{resultados}

The use of local squeezing operations enlarges the set of setups available to Alice or Bob for non-locality tests. In previous work the setups were determined by the choices of displacements pairs. The generalization introduced in this work allows the setups to be chosen within an enlarged set of combinations of squeezing and displacement.\\

An interesting possibility, not previously explored, results from the  use of setups involving local squeezing operations and no displacements. Specifically, we refer to the case where Alice and Bob consider two setups corresponding to squeezings parameters  $\{u,\ u^{\prime}\}$ and  $\{v,\ v^{\prime}\}$ respectively while keeping $\alpha=\beta=0$. 

In this case the SEQPD are evaluated at the phase space origin. Consequently their value are entirely determined by the corresponding covariance matrix determinant [see Eq. \eqref{Wigner_de_dos_modos}]. The CHSH inequality for $-1 \leq s \leq 0$ becomes \cite{Lee09}:
\begin{align}\label{BellLee2}
&\vert(1-s)^4\mathbb{D}(u,u^{\prime},v,v^{\prime},s)\nonumber\\
&+2 s(1-s)^2\left[ \dfrac{1}{\sqrt{(pu+\vert s\vert)(p/u+\vert s\vert)}}\right.\nonumber \\
&\left. +\dfrac{1}{\sqrt{(qv+\vert s\vert)(q/v+\vert s\vert)}}\right]+2s^2\vert \leq 2
\end{align}
with
\begin{eqnarray}\label{Duv}
&& \mathbb{D}(u,u^{\prime},v,v^{\prime},s) = (\det\sigma_{uvs})^{-\frac{1}{2}}+(\det\sigma_{u^{\prime}vs})^{-\frac{1}{2}}\nonumber\\ &+&(\det\sigma_{uv^{\prime}s})^{-\frac{1}{2}}-(\det\sigma_{u^{\prime}v^{\prime}s})^{-\frac{1}{2}} 
\end{eqnarray}
where $\sigma_{uvs}\equiv \mathcal{S}_{uv}\sigma \mathcal{S}_{uv}^{T} + \vert s\vert\mathds{1}_4$ (see the 
Apendix for details).\\

We first notice that for $s=0$ (corresponding to measurements of the parity operator) all four determinants appearing in \eqref{Duv} are equal which is a consequence of \eqref{substituciondesigmaconsqueezing} and the fact that the covariance matrix determinant is unchanged by symplectic operations \cite{Adesso04}. In this case the inequality \eqref{BellLee2} reduces to: $2\mu \leq 2$ where $\mu=1/\sqrt{\det \sigma}$ is the global state purity. The inequality is always fulfilled and saturated for pure states. This result shows that BNL cannot be proven through local squeezing followed by the measurement of the (non-displaced) parity observable. 

However, for $s < 0$, the inequality \eqref{BellLee2} can be violated. In the following we concentrate in the $s=-1$ case for which the setups required by Alice and Bob to reveal BNL reduce, on each system, to a squeezer (implementing up to two different quadrature compressions) and a detector able to discriminate light from darkness (Fig. \ref{setup}). \\

In principle, maximization of the l.h.s. of inequality \eqref{BellLee2} require the variation of all four compressions factors $u,u^{\prime},v,v^{\prime}$. In analogy with the BW maximization scheme, we have also considered the case where only one of the compression factors used by each part is different from one.\\

\begin{figure}[ht!]
\centering
\includegraphics[width=.8\linewidth]{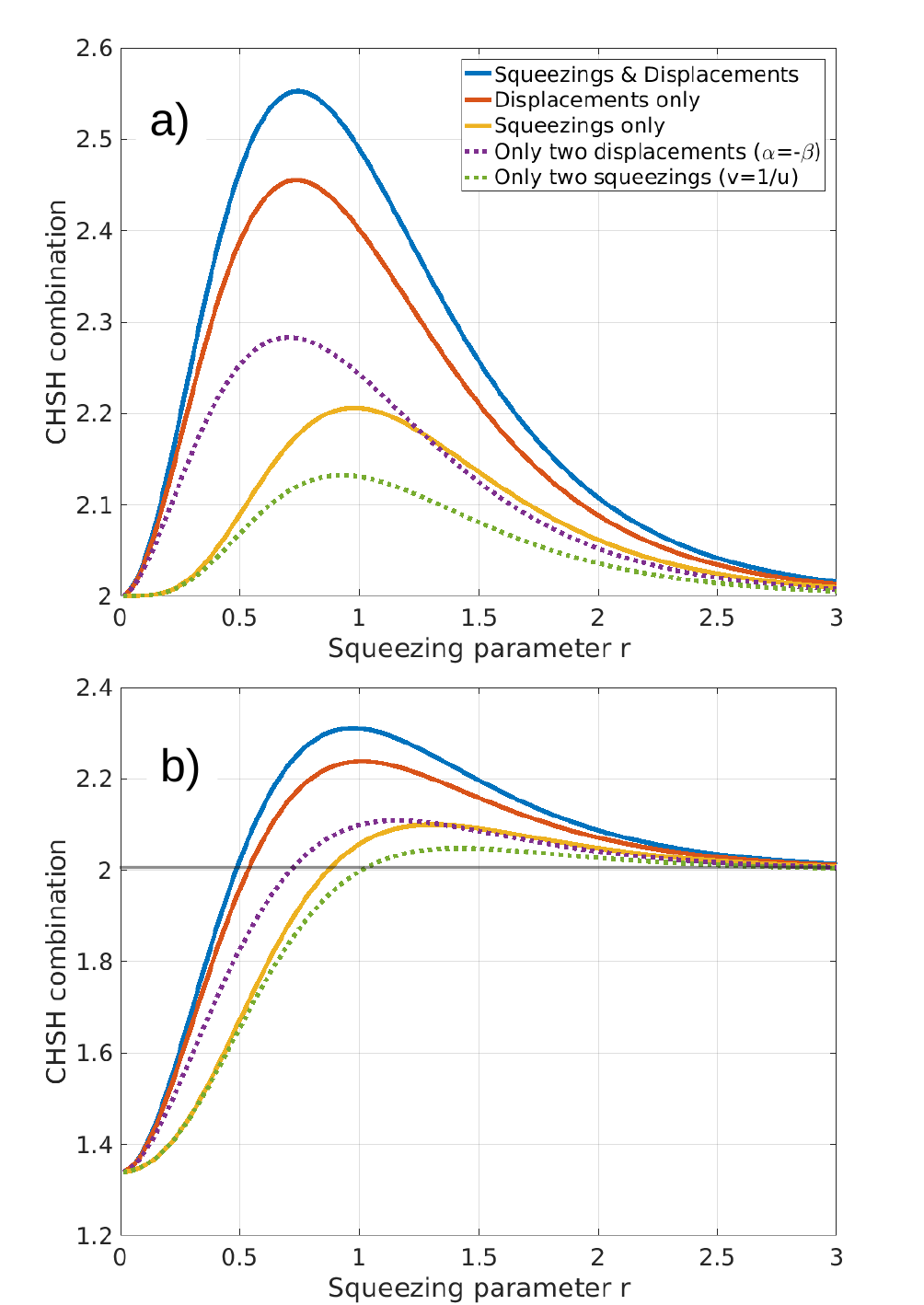}
\caption{\label{todo} Evaluation of the maximum value of $\mathcal{B}_{CHSH}$ of the CHSH combination in inequality  \eqref{CHSHChi} for $s=-1$ as a function of the squeezing parameter $r$ for a pure TMSV state (a) and a symmetrical TMSTS with global purity $0.7$ (b). Solid blue: all compression factors and displacements are freely varied. Solid red: only the displacements are freely varied with no squeezing. Dashed violet: BW maximization scheme (only one nonzero displacement on each part), no squeezing. Solid yellow: Freely varied squeezings with no displacements. Dashed green: only one compression factor different from one on each part, no displacements. Bell non-locality is proven for $\mathcal{B}_{CHSH} > 2$} 
\end{figure} 

We have used several maximization strategies for $\mathcal{B}_{CHSH}$ in \eqref{CHSHChi} for specific examples of two-mode Gaussian states. We stress, nonetheless, that the BNL test presented above can be applied to any two-mode Gaussian state. 
\\

The two examples presented in Fig. \ref{todo} belong to the general category of two-mode squeezed thermal states (TMSTS) with squeezing parameter $r$ for which the coefficients of the covariance matrix in its standard form are given by: $ p=\nu_1\cosh^2(r)+\nu_2\sinh^2(r),\ q=\nu_1\sinh^2(r)+\nu_2\cosh^2(r), \ m=-n=(\nu_1+\nu_2)\sinh(2r)/2$ with $\nu_1 \geq 1$ and $\nu_2 \geq 1$. Symmetric TMSTS have $\nu_1= \nu_2$. The case $\nu_1= \nu_2=1$ is a pure TMSV state. \\
 
Figure \ref{todo}a) refers to pure TMSV states. The plots represent the maximum value of $\mathcal{B}_{CHSH}$  in inequality  \eqref{CHSHChi} for $s=-1$ and different maximization schemes as a function of the state squeezing parameter $r$ (not to be confused with the  local squeezing operations parameters). Bell inequality violations occur for all values of $r$. The solid-blue line corresponds to the case where all four local squeezings and four displacements are freely varied. The largest Bell inequality violation $\mathcal{B}_{CHSH}=2.55$ occurs at $r=0.75$ for displacements $\alpha=-\beta=-0.17$, $\alpha^{\prime}=-\beta^{\prime}=0.62$ and compression factors $u=v=0.97,\ u^{\prime}=v^{\prime}=1.65$. Such value represents a substantial improvement with respect to the maximum violation $\mathcal{B}_{CHSH}=2.45\ (r=0.74)$ \cite{Lee09} obtained for $\alpha=-\beta=-0.15$, $\alpha^{\prime}=-\beta^{\prime}=0.52$ when only the displacements are freely varied with no squeezing (solid-red line). The dashed-violet line correspond to the BW maximization scheme were only one of the displacements realized by the two parts is nonzero \cite{Banaszek98}). $\mathcal{B}_{CHSH}=2.28$ is achieved in this case at $r=0.70$ for $\alpha=-\beta=0.86$. \\

The use of squeezing operations not only can improve the Bell inequality violations obtained by the use of displacements only. It can also suffice to reveal BNL. This is shown by the solid-yellow and dashed-green plots in Fig. \ref{todo}a). The solid-yellow line correspond to the case where all four local squeezings are freely varied. A maximal $\mathcal{B}_{CHSH}=2.21$  is obtained at $r = 0.98$ for 
compression factors $u=1/v=1.41,\ v^{\prime}=1/u^{\prime}=3.67$. In fact, just one local squeezing operation on each side is sufficient to reveal BNL as shown by the dashed-green line. A maximum $\mathcal{B}_{CHSH}=2.13$ is obtained at $r=0.93$ for compression factors $u=1/v=3.2,\ u^{\prime}=v^{\prime}=1$.\\

Figure \ref{todo}b) presents the results for a mixed, symmetrical TMSTS with $\nu_1=\nu_2=1.2$ (global purity $\mu=0.7$). In this case, Bell inequality violations only occur for sufficiently large $r \gtrsim 0.5$. Notice that for $r\simeq 0.5$ observables relying only on displacement do not violate the CHSH inequality while BNL is reveled by observables involving squeezing in addition to displacement.\\

The results presented so far show that the possibility to identify BNL two-mode Gaussian states is increased by the use of local squeezing operations. A natural concern refers to the extend by which such advantage can be affected by the use of non ideal detectors with efficiency $\eta < 1$. This problem can be readily addressed along the same lines as in \citep{Lee09}. It is a consquence of the formal similarity between the definitions of the observables $\hat{\Pi}(\xi;\alpha;s)$  in \eqref{nuevoPi} and the observables $\hat{\Pi}(\alpha;s)$ in \citep{Lee09} that the same treatment of the finite detector efficiency can be applied. It amounts to replacing the parameter $s$ in the SEQPDs $W(\chi_a,\chi_b,s),\ W(\chi_a,s)$ and $ W(\chi_b,s)$ appearing in inequalities \eqref{CHSHChi} by $s^{\prime}=-(1-s-\eta)/\eta$ \citep{Lee09}.

\begin{figure}[ht!]
\centering
\includegraphics[width=1\linewidth]{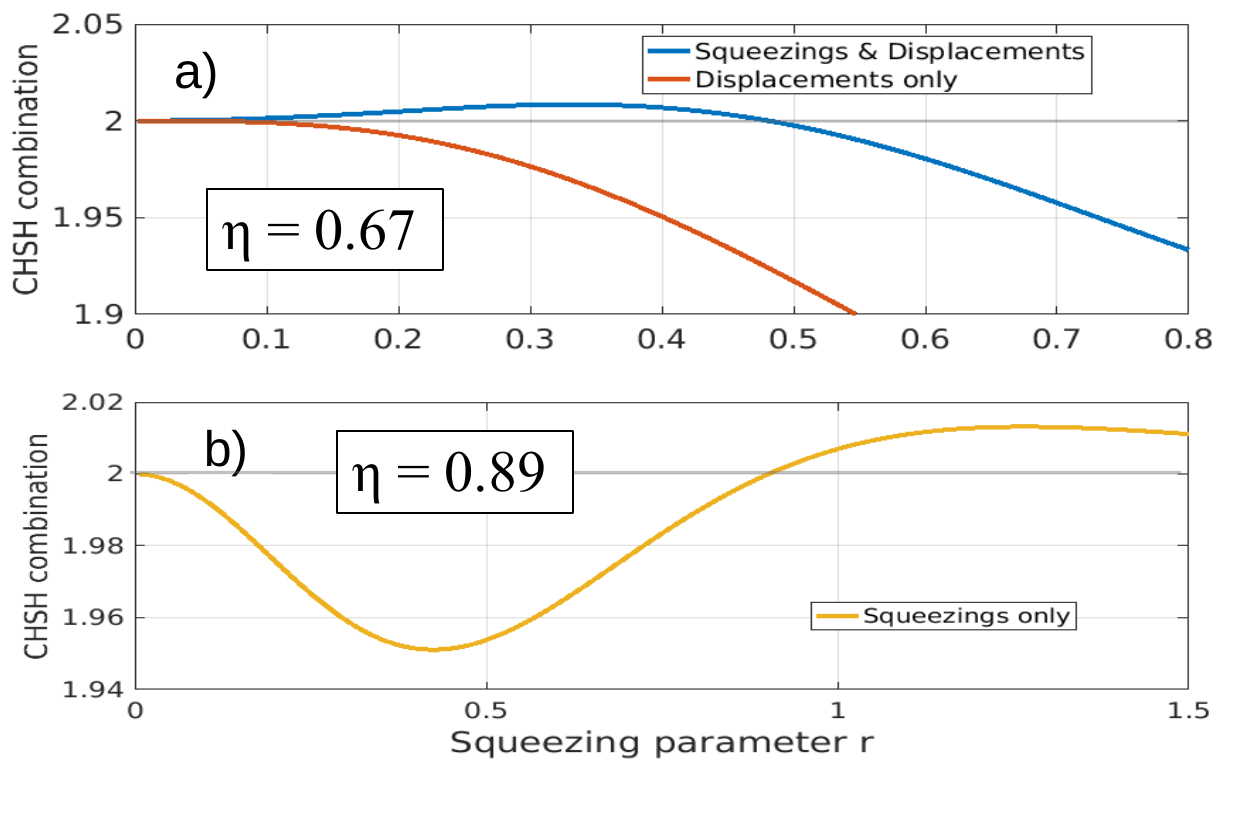}
\caption{\label{eta} Evaluation of the maximum value of $\mathcal{B}_{CHSH}$ of the CHSH combination in inequality  \eqref{CHSHChi} for $s=-1$  for a pure TMSV as a function of the state squeezing parameter $r$. a) Common detector efficiency: $\eta = 0.67$. BNL cannot be proven at $r=0.4$ if  observables involving only displacements are used (red line). It can be revealed by observables using combined displacements and squeezing (blue line). b) Common detection efficiency $\eta = 0.89$. BNL can be proven for $r\gtrsim 0.9$ by observables using only local squeezing operations.} 
\end{figure} 

Figure \ref{eta} shows two examples of the performance of the extended family of observables for the identification of BNL with inefficient detectors. The same overall detection efficiency $\eta$ is assumed for Alice and Bob. In both cases the inequality \eqref{CHSHChi} is evaluated for $s=-1$ which only requires a detector discriminating light from no-light.\\ 

Figure \ref{eta}a) illustrates the increased tolerance to detection inefficiency resulting from the use of observables including squeezing operations. It is mentioned in \citep{Lee09} that for $s=-1$ a minimum efficiency $\eta \gtrsim 0.75$ is required to reveal BNL in a TMSV state with $r=0.4$ using observables involving only displacements. As shown in Fig. \ref{eta}a) BNL can be revealed for such state with a detection efficiency as low as $\eta \simeq 0.67$ using observables involving local squeezing operations in combination with displacements. [$\alpha=-\beta=-0.13$, $\alpha^{\prime}=-\beta^{\prime}=0.62$ and compression factors $u=v=1.05,\ u^{\prime}=v^{\prime}=1.65$]\\

If the considered observables only involve squeezing operations and no displacements, BNL can still be revealed for overall detection efficiencies  $\eta > 0.88$ for TMSV states with $r \gtrsim 1$ as illustrated in Fig. \ref{eta}b).\\

\section{Conclusions}\label{sec:conclusiones}

Continuous variables states and Gaussian states in particular, are a useful resource for quantum information and quantum communications protocols \cite{Braunstein05,Adesso14}. Most protocols rely on the sharing of an entangled state between two parties. The observation of Bell non-locality is a powerful method for entanglement certification \cite{Brunner14,Chen21}.\\ 

In this work we have presente an extended  set of local observables involving  squeezing operations that can be used to reveal Bell non-locality in two-mode Gaussian states through violations of CHSH inequalities. The extended set of observables can lead to an increase of the maximum CHSH inequality violation with respect to previously considered  observables resulting in enhanced non-locality sensitivity. \\

The extended set includes observables that do not require displacements and may suffice to reveal BNL in some two-mode Gaussian states. The corresponding experimental scheme, required by each part, is conceptually very simple. It reduces (for $s=-1$) to a single-mode squeezer followed by a light detector only required to discriminate between vacuum and no vacuum states. Present day photodetectors can achieve this with large efficiency in the visible and near IR. The squeezing operation can be realized through degenerate parametric down-conversion or four-wave mixing \cite{Gerry04,Schnabel17}. The required amount of squeezing, less than 6 dB in the examples considered, is well within state of the art \cite{Vahlbruch16,Schonbeck18}.  

\section{Acknowledgments}
This work was supported by ANII, CSIC and PEDECIBA (Uruguayan agencies).\\

\appendix*
\section{Evaluation of the squeezing-extended quasi-probability distributions}
For a two-mode system the Wigner function of a Gaussian state (ignoring displacements) is given by:
\begin{eqnarray}\label{Wigner_de_dos_modos}
W(\alpha,\beta) &=& \frac{4}{\pi^{2}\sqrt{\det \sigma}}e^{-\frac{1}{2}X^{T}\sigma^{-1}X}
\end{eqnarray}
where $X^{T}$ is the row vector $[\alpha+\alpha^{*}, -i(\alpha-\alpha^{*}), \beta+\beta^{*}, -i(\beta-\beta^{*})]$ and $\sigma$ is the $4\ \times\ 4$ covariance matrix.\\ 

It is well known that for non-positive $s$ the sQPD can be obtained from the Wigner function by convolution \cite{Leonhardt10}. For a two-mode Gaussian state we have:
\begin{eqnarray}
W(\alpha,\beta,s) &=& \frac{1}{4s^2 \pi^2}\int d^4Y W(Y) e^{-\frac{1}{2}(Y-X)^{T}\vert s \vert^{-1}(Y-X)}\nonumber\\
\label{Ws}
\end{eqnarray} 

Using well known properties of Gaussian convolution, the sQPD can be obtained from the Wigner function \eqref{Wigner_de_dos_modos} through the substitution:
\begin{eqnarray}
\sigma \rightarrow \sigma_s \equiv \sigma + \vert s\vert\mathds{1}_4 \label{substituciondesigma}
\end{eqnarray}
where $\mathds{1}_4$ is a 4 by 4 unit matrix.\\

Generalization of the previous result to SEQPD is straightforward. It is obtained by the the following replacement in \eqref{Wigner_de_dos_modos}:
\begin{eqnarray}
\sigma \rightarrow \sigma_{Es} \equiv \mathcal{S}\sigma \mathcal{S}^{T}+ \vert s\vert\mathds{1}_4 \label{substituciondesigmaconsqueezing}
\end{eqnarray}
where $\mathcal{S}$ is the symplectic transform matrix describing the local squeezing operation \cite{Adesso14}.\\ 

It is important to remark that for $s\neq 0$ the SEQPD $W(\xi;\alpha;s)$ obtained from \eqref{Wigner_de_dos_modos} by the replacement \eqref{substituciondesigmaconsqueezing} is generally not the result of a variable change in phase-space. More precisely, the SEQPD is \emph{not equal} to the corresponding sQPD $W(\alpha;s)$  evaluated at the canonically transformed phase-space position  $X^{\prime} = \mathcal{S}^{T}X$. This is because for squeezing operations:
\begin{eqnarray}
\mathcal{S}\sigma \mathcal{S}^{T}+ \vert s \vert \mathds{1} &\neq & \mathcal{S}(\sigma+ \vert s \vert \mathds{1})\mathcal{S}^{T}
\end{eqnarray}
since $\mathcal{S}\mathcal{S}^{T} \neq \mathds{1}$. Also,
\begin{eqnarray}
det{\sigma_{Es}} &\neq & det{\sigma_{s}} 
\end{eqnarray}

An important exception to the previous remark is the Wigner function for which
\begin{eqnarray}
W(\xi;X;s=0) = W(\mathcal{S}^{T}X)\label{solowigner}
\end{eqnarray}.
In consequence for $s=0$ the same information can be obtained from the sampling of the Wigner function or the SEQPD.\\ 

Similarly, no information can be gained by the local application of a phase-space rotation (instead of squeezing) since in such case the symplectic transform matrix is orthogonal ($\mathcal{S}\mathcal{S}^{T} = \mathds{1}$).

%

\end{document}